\documentclass[aps,pre,twocolumn,floatfix,showpacs,amsmath,10pt]{revtex4-1}
\usepackage{amssymb,graphics}
\usepackage{setspace} 
\usepackage{graphicx}
\usepackage{hyperref}
\usepackage{xcolor}

\newcommand{\showfigures}[1]{{#1}} 

\newcommand{\red}[1]{\textcolor{black}{{#1}}}
\newcommand{\DP}[2]{\ensuremath{\frac{\partial{#1}}{\partial{#2}}}}
\newcommand{\DPn}[3]{\ensuremath{\frac{\partial^{#1}{#2}}{\partial{#3^{#1}}}}}
\newcommand{\DT}[2]{\ensuremath{\frac{\mathrm{d}{#1}}{\mathrm{d}{#2}}}}

\newcommand{\kd}{k_{\mathrm{d}}}
\newcommand{\td}{\tau_{\mathrm{d}}}

\newcommand{\rhoh}{\rho_{\mathrm{d}}}
\newcommand{\rhoe}{\rho_{\mathrm{e}}}

\newcommand{\se}{s_{\mathrm{e}}}
\newcommand{\sebar}{\bar{s}_{\mathrm{e}}}
\newcommand{\kbar}{\bar{k}_{\textrm{d}}}

\newcommand{\sigp}{\sigma_{\mathrm{p}}}
\newcommand{\sige}{\sigma_{\mathrm{e}}}
\newcommand{\sigh}{\sigma_{\mathrm{h}}}
\newcommand{\siga}{\sigma_{\mathrm{A}}}

\newcommand{\etascale}{\tilde{\eta}}

\newcommand{\del}{\partial}

\renewcommand{\(}{\left(}
\renewcommand{\)}{\right)}

\newcommand{\param}{\alpha}

\begin{document}
\title{One-dimensional collective migration of a proliferating cell monolayer}
\author{Pierre Recho$^{1,3}$}
\author{Jonas Ranft$^2$} 
\author{Philippe Marcq$^3$}
\email{philippe.marcq@curie.fr}

\affiliation{$^1$ Mathematical Institute, University of Oxford,
Oxford OX26GG, United Kingdom}
\affiliation{$^2$ Laboratoire de Physique Statistique, 
\'Ecole Normale Sup\'erieure, 24 rue Lhomond, F-75231 Paris Cedex 05, France}
\affiliation{$^3$ Laboratoire Physico-Chimie Curie,
Institut Curie, Universit\'e Marie et Pierre Curie, CNRS UMR 168, 
26 rue d'Ulm, F-75248 Paris Cedex 05, France}

\date{January 4, 2016}

\begin{abstract}
The importance of collective cellular migration during embryogenesis 
and tissue repair asks for a sound understanding of underlying principles 
and mechanisms. Here, we address recent \emph{in vitro} experiments 
on cell monolayers which show that the advancement of the leading edge 
relies on cell proliferation and protrusive activity at the tissue margin. 
Within a simple viscoelastic mechanical model amenable to detailed analysis,
we identify a key parameter responsible for tissue expansion, and we 
determine the dependence of the monolayer velocity as a function of 
measurable rheological parameters. Our results allow us to discuss the 
effects of pharmacological perturbations on the observed tissue dynamics.
\end{abstract}

\maketitle

\section{Introduction}
\label{sec:intro}

Recent experiments on the expansion dynamics of epithelial cell monolayers 
highlighted a propagative mode with an approximately constant velocity 
at the leading edge. They were performed with Madin-Darby canine kidney 
(MDCK) epithelial cells in a quasi-one-dimensional geometry, 
first on tracks of small width \cite{Vedula2012},
then on a (cylindrical) fiber of small radius \cite{Yevick2015}.
The presence of a free boundary gives rise to an inhomogeneous cell density 
along the tissue, increasing monotonically towards the rear of the cell layer.
Conversely, the velocity is maximal at the leading edge
and decreases monotonically with the distance from the front. 
Collective cell migration in the rear 
comes to a halt as cell density rises with time \cite{Yevick2014}. 

Studying epithelization over durations short
compared to the typical cell cycle, some among us showed that a cell 
monolayer on a hard substrate may be described by an inviscid,
incompressible fluid driven by active boundary forces 
\cite{Cochet-Escartin2014}. A detailed comparison of model predictions with 
experimental measurements showed that during the epithelization of
a disc-shaped empty domain, protrusive activity at 
the leading edge dominated force generation and external 
friction between monolayer and substrate dominated energy dissipation.

Motivated by the experiments mentioned above, we wish to extend these 
results to monolayer expansion assays whose 
duration is longer than a typical cell cycle. 
In addition to lamellipodial activity at the leading edge,
cell proliferation, as well as inflow of cells from the reservoir, 
may drive collective migration. We take into account variations 
of the cell density and use linear non-equilibrium thermodynamics to relate
the equation of state of the tissue to the dependence of the proliferation
rate on cell density. 
In doing so, we aim to provide a biophysical understanding of the 
tissue dynamics observed in experiments, which further allows to account 
for the effects of drug treatments that modify the velocity at the 
leading edge~\cite{Vedula2012,Yevick2015}. Our approach is complementary 
to earlier studies addressing mainly the kinetics  of collective migration 
without reference to the tissue mechanical behaviour 
\cite{Murray2002,Maini2004a,Maini2004b}.
In the following, the cell monolayer will be loosely referred to as 
the ``tissue'' although it lacks some of the complexity of \emph{in vivo} 
tissues, such as, \emph{e.g.}, collagen secretion and organization.

This article is organized as follows. 
In Sec.~\ref{sec:model}, we introduce a one-dimensional mechanical description 
of the collective migration of a proliferating cell monolayer.
The combination of forces generated by protrusions at the leading edge
and by proliferation in the bulk leads to monolayer expansion 
at constant front velocity. We thus study in Sec.~\ref{sec:traveling} 
the existence of traveling wave solutions and the dependence of 
the front velocity upon control parameters.  
The limit of vanishing viscosity, presented in 
Sec.~\ref{sec:inviscid}, gives rise to an effective Fisher-Kolmogorov-like 
free-boundary problem which has a clear mechanical interpretation. 
In Sec.~\ref{sec:perturb}, we discuss how pharmacological perturbations 
may modify the response of the tissue.
Finally, we summarize and discuss our results in Sec.~\ref{sec:conc}.

\section{Expansion of a proliferating cell monolayer}
\label{sec:model}

With highly cohesive cell monolayers in mind, we formulate a continuum
description of collective migration. We describe the tissue at length 
scales large compared to the size of a single cell, and thus define 
the coarse-grained fields of the cell number density $\rho(\vec{r},t)$, 
the velocity $v(\vec{r},t)$ and the stress $\sigma(\vec{r},t)$ as 
functions of the space coordinate $\vec{r}$ and time $t$. 
Sec.~\ref{sec:model:basis} recapitulates the conservation laws obeyed by
the system. Sec.\ref{sec:model:thermodynamics} shows that
the framework of linear non-equilibrium thermodynamics relates
the cell density dependence of stress and proliferation rate.
Sec.~\ref{sec:model:model} gives the resulting evolution equations 
and boundary conditions in dimensionless form.

\subsection{Conservation laws}
\label{sec:model:basis}

In order to describe the migration of a cell monolayer along narrow tracks 
or along cylindrical fibers of small radii, we assume that the relevant 
fields vary little across the track width or the fiber circumference. 
\red{This assumption implies the absence of long-range velocity correlations
that are known to occur for track widths larger than 
$~100\,\mu{\rm m}$~\cite{Vedula2012}, the order of magnitude of 
the velocity correlation length as measured in bulk.}

\begin{figure}[!t] 
\begin{center}
\showfigures{
\includegraphics[scale=0.9]{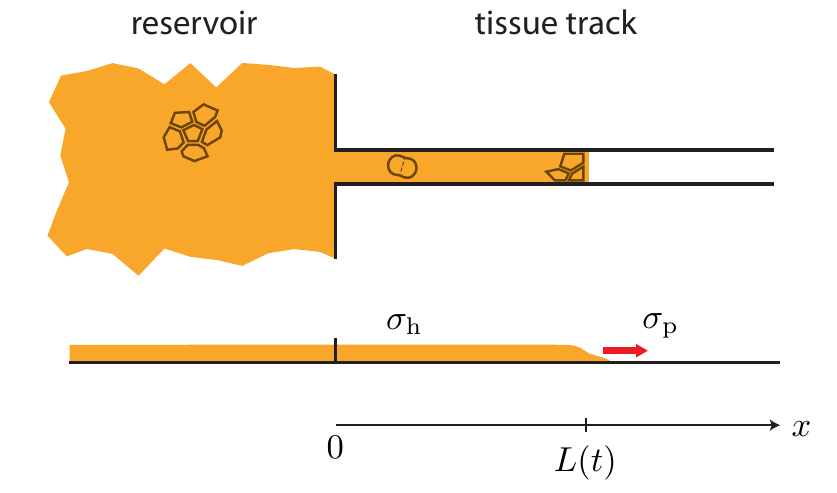}
}
\caption{
\label{fig:sketch} 
\textbf{Schematic representation of the one-dimensional expansion of
a proliferating cell monolayer.} (Top) A reservoir, which is 
occupied by a confluent cell monolayer, is connected to a smaller track
to which the advancing cells are confined. Cells can undergo division
and apoptosis, and cells at the leading edge exert pulling forces 
on the tissue due to lamellipodial activity. (Bottom) Side view of 
the upper sketch; $\sigh$ denotes the homeostatic stress at which cell 
division and cell death balance, and $\sigp$ denotes the protrusive stress 
exerted at the leading edge. (Online version in colour.)
}
\end{center}
\end{figure}

Within a thin film approximation, we consider an effectively 
one-dimensional system along the migration axis: $\vec{r} = x \, \vec{e}_x$.
At time $t$, the tissue covers the track, or the fiber,
from a reservoir at $x=0$ to the leading edge at $L(t)$,
see Fig.~\ref{fig:sketch} for a sketch.
Importantly, cells belong to a monolayer everywhere, including in the 
reservoir, and there is no ``permeation''
of cells from superior layers
as in the case of a cell  monolayer spreading from an aggregate with 
weaker cell-cell adhesion \cite{Douezan2012}. 
Since the cell monolayers that we consider are easily supplied with nutrients 
from the third dimension, \emph{i.e.}, from the culture medium at 
the apical side, growth is not limited by nutrient diffusion \red{within the tissue. 
Because the extracellular fluid is not confined to the tissue layer, we can 
neglect interstitial flows that would otherwise give rise to additional mechanical 
constraints~\cite{Byrne2003,Ranft2012}.
We furthermore do not consider here the effect of external chemotactic fields \cite{BenAmar2014}.}

\paragraph*{Cell conservation law.} 
The cell number density obeys \red{the mass balance equation, supplemented} 
with a source term due to cell proliferation:
\begin{equation}
\label{eq:cons:matter}
\del_t \rho + \del_x \(\rho v\) = \kd(\rho) \, \rho 
\end{equation}
where  $\kd$ is the effective proliferation rate, combining
cell divisions and cell deaths (delaminations), and depends on 
cell density, see below.
This conservation law is associated with boundary conditions 
at the inlet, where the tissue is connected to the reservoir, 
and the free end of the tissue. In line with experiments, we 
impose zero flux boundary conditions: 
\begin{align}
\label{eq:BC:0}
v(0,t) &= 0 & \text{and} \\
\label{eq:BC:L:v}
v(L(t),t) &= \dot L(t) \,,
\end{align}
respectively. Throughout the text, $\dot{}=\mathrm{d}/\mathrm{d}t$ denotes the 
total derivative with respect to time. Eq.~\eqref{eq:BC:0} assumes 
that cell proliferation in the reservoir and subsequent tissue inflow
is negligible compared to proliferation in the tissue, 
as observed experimentally \cite{Yevick2014}. 
Eq.~\eqref{eq:BC:L:v} provides a kinematic condition for the evolution of 
the leading edge. In principle, it could be modified by including a boundary growth term
if necessary.

\paragraph*{Proliferation rate.}
Consistent with observations \cite{Puliafito2012,Yevick2014}, 
we consider that the proliferation rate depends on cell density and 
assume for simplicity a linear relation \cite{Basan2009,Ranft2010},
\begin{equation}
\label{homeogrowth}
\kd(\rho) = \frac{1}{\td} \frac{\rhoh - \rho}{\rhoh} \,,
\end{equation}
where $\rhoh$ is the tissue carrying capacity, in other words the reference 
cell density at which the net cell division rate vanishes. 
[See Appendix~\ref{sec:app:gen} for the more general case of
arbitrary $\kd(\rho)$.]
The characteristic time scale $\td$ may be estimated 
experimentally \cite{Puliafito2012}
\red{for} vanishing cell densities as \red{$\td = \kd(0)^{-1}$}.
Note that the coupling between density and proliferation is not
\emph{per se} mechanical but may be a combined effect of mechanical
compression and density-dependent signaling between cells. One declared
aim of this paper is to investigate the ramifications of such a coupling
for migrating tissues.

\paragraph*{Force balance.} 
In this continuum framework, internal and
external forces are respectively described by the stress field $\sigma(x,t)$
and the external force field $f^{\rm ext}(x,t)$.
Force balance is expressed as
\begin{equation}
\label{eq:fbal}
\del_x \sigma = -f^{\rm ext} \,.
\end{equation}
We assume that external forces are due to fluid friction
between the tissue and the substrate \cite{Cochet-Escartin2014}:
\begin{equation}
\label{eq:friction}
f^{\rm ext} = - \xi v  \,.
\end{equation}
For simplicity, we neglect active, bulk motility forces, as may be produced by,
\emph{e.g.}, cryptic lamellipodia within the monolayer \cite{Farooqui2005}.
At the free boundary $x=L(t)$ however, \red{leading cells extend 
lamellipodia and exert active pulling forces on the rest of the tissue. 
Mechanically, we therefore treat boundary cells at the front 
as external agents applying a tensile traction $\sigma_{\text{p}}$
on the monolayer,}
leading to the following condition on the stress \cite{Cochet-Escartin2014}:
\begin{equation}
\label{eq:BC:sigma}
\sigma(L(t),t)=\sigp \,.
\end{equation}

\subsection{Linear non-equilibrium thermodynamics}
\label{sec:model:thermodynamics}

In order to close the system of equations and to fully specify the tissue 
mechanics, one needs a constitutive equation to relate stresses to 
quantities like the cell density and the cell velocity field. 
It is generally assumed that an equation of state relates 
the isotropic tissue stress to the cell density, and that 
viscous stresses occur in the presence of velocity gradients, 
effectively leading to a Kelvin-Voigt viscoelastic rheology
in one dimension.

Instead of postulating \emph{ad hoc} such a relation, 
we use the framework of linear non-equilibrium thermodynamics 
\cite{Chaikin2000} to specify the constitutive equation of the cell 
monolayer, see \cite{Recho2014} for a similar derivation in the context of 
single cell migration. This approach has the advantage of being systematic 
in terms of the identified fluxes and forces once a free energy is specified, 
and to give a coherent picture of the dissipation caused by the system. 
More specifically, it allows us to identify an equation of state for the 
tissue stress that is thermodynamically consistent with the 
linear relation $\kd(\rho)$ chosen above. 
\red{To express the dissipation rate in the material, we follow the classical 
Coleman-Noll method \cite{Truesdell2004, Gurtin2010, Berdichevsky2009} 
(see also \cite{Epstein2000} for a general treatment in the context of 
volumetric growth).}

However, as one may argue that tissues are too far from (thermodynamic) 
equilibrium for linear non-equilibrium thermodynamics to hold and that 
the assumption of the existence of a well-defined free-energy density 
that depends on a few variables of the system may not be justified, 
we refer the reader to Appendix~\ref{sec:app:gen} for a more general treatment. 

In the presence of distributed (bulk) friction forces  $f^{\rm ext} = -\xi v$ 
and of (boundary) traction forces $\sigp$ due to lamellipodial protrusions 
at the leading edge, the power $\Pi$ of the \emph{external} forces to 
which the monolayer is subjected reads
\begin{align*}
\Pi &= -\int_{0}^{L}\xi v^2 \, \mathrm{d}x+\sigp \dot{L} \\
&= \int_{0}^{L}(-\xi v^2+\partial_x(\sigma v)) \, \mathrm{d}x \,.
\end{align*}
By taking into account the force balance equation~\eqref{eq:fbal},
$\Pi$  can also be expressed as the power of the \emph{internal} forces
\begin{equation*}
\Pi= \int_{0}^{L}\sigma\partial_xv \, \mathrm{d}x.
\end{equation*}
The total free energy of the cell monolayer reads
\begin{equation*}
F=\int_{0}^{L} \rho \, f(\rho) \, \mathrm{d}x,
\end{equation*}
where $f(\rho)$ is the specific free energy, which we assume to be dependent  
only on the cell density $\rho$ since temperature is constant in the tissue and 
heat fluxes can thus be neglected. A theory describing a more general growth 
process is currently under investigation and will be published elsewhere. 
Using Reynolds' theorem, the rate of change of the free energy reads
\begin{equation*}
\dot{F} = \int_{0}^{L} \kd \rho f \, \mathrm{d}x +
\int_{0}^{L} \rho^2\frac{\mathrm{d}f}{\mathrm{d}\rho}
(\kd-\partial_xv) \,\mathrm{d}x.
\end{equation*}
In this isothermal system the power of external forces and 
the free energy rate must satisfy a dissipation principle
\begin{equation*}
\Sigma = \Pi-\dot{F}\geq 0.
\end{equation*}
We express $\Sigma$ as a bilinear form
\begin{equation*}
\Sigma = \int_{0}^{L} (\sigma+p)
\partial_xv \, \mathrm{d}x +
\int_{0}^{L} (-\rho\mu)
\kd \, \mathrm{d}x.
\end{equation*}
\red{where $p=\rho^2 \frac{\mathrm{d}f}{\mathrm{d}\rho}$
is the thermodynamic pressure and 
$\mu=f+\frac{p}{\rho}$
is the thermodynamic chemical potential \cite{Lowengrub1998} whose 
role in three dimensional elasticity is played by the corresponding 
component of the energy momentum Eshelby tensor \cite{Epstein2000}. 
The two terms under integrals can be interpreted as  products of the 
thermodynamic forces $\sigma + p$, $-\rho\mu$ with the respective conjugate 
thermodynamic fluxes $\partial_x v$, $\kd$.}

Constitutive equations are obtained by expressing 
thermodynamic forces as a linear combination of thermodynamic forces
through Onsager type relations. \red{For simplicity, we neglect 
the cross-terms, and obtain}  
\begin{subequations}
\label{forceflux}
\begin{align}
\sigma+p&=l_{11}\,\partial_x v \\
-\rho \mu &=l_{22} \, \kd. 
\end{align}
\end{subequations}
Here the different tensorial nature of the fluxes/forces is not an 
issue because of the 1D Ansatz. Both diagonal kinetic 
coefficients $l_{11}$ and $l_{22}$ are positive and we shall from now on
denote $\eta=l_{11}$ the viscosity.  
We are left with the conventional relation (\ref{forceflux}a) for a 
viscous, one-component, compressible fluid 
at a constant temperature \cite{Lowengrub1998}
\begin{equation*}
\sigma=-p+\eta\partial_xv
\end{equation*}
Identity (\ref{forceflux}b) relates the proliferation rate to its 
naturally associated generalized force, the chemical potential 
\cite{Garikipati2004,Ambrosi2010}.
Interestingly, the only choice of free energy consistent 
with the form of $\kd$ assumed in Eq. \eqref{homeogrowth} leaving $\rho$ 
unconstrained, is then 
\begin{equation*}
f(\rho) = \frac{l_{22}}{\rho \, \td}
\left( \log \left(\frac{\rhoe}{\rho}\right) + \frac{\rho}{\rhoh} - 1 \right)
\, ,
\end{equation*}
where we have imposed the condition \red{$p(\rhoe)=0$}, introducing an 
\emph{elastic} reference density $\rhoe$. Importantly, $\rhoh \neq \rhoe$: 
the carrying capacity does not need to be equal to the elastic reference 
density. This leads to an expression of the stress thermodynamically 
consistent with \eqref{homeogrowth},
\begin{equation}
  \label{eq:def:stress:visc}
  \sigma = - E \, \log\left( \frac{\rho}{\rhoe} \right)  + \eta \, \DP{v}{x} \, , %
\end{equation}
where we identify the prefactor $E=l_{22}/\td$ as the elastic 
modulus of the tissue. 

The monotonically decreasing function $\se(\rho) = -\log(\rho/\rhoe)$ 
characterizes the dependence of (elastic) stress upon density allowing to
consider large deformations of the material. 
It is identical to the true strain, and infinitely penalizes both infinite 
dilution ($\rho=0$) and the formation of 
singularities ($\rho=\infty$). While the same form was postulated 
for convenient technical reasons in \cite{Mi2007,Arciero2011,Stepien2014} 
to describe wound healing and cell colony expansion, we show here  that it is 
consistent with a proliferation rate linear in the density as assumed
in Eq.~\eqref{homeogrowth}.

\subsection{The model}
\label{sec:model:model}

We now formulate the problem of cell monolayer expansion,
assuming for simplicity that the material parameters 
$\td$, $\rhoh$, $\eta$, $\xi$, $\sigp$, $\rhoe$ and $E$ 
are constant.

\red{We use} $\td$, $\sqrt{\frac{E \, \td}{\xi}}$,  
$\rhoh$ and $E$ as units of time, length, 
density and stress respectively, and the reduced stress field 
\begin{equation*}
  \label{eq:reduced:stress}
s= \frac{\sigma-\sigp}{E}.   
\end{equation*}
\red{The velocity field can be expressed as a stress gradient using 
\eqref{eq:fbal} and \eqref{eq:friction}. Combining 
\eqref{eq:cons:matter} with \eqref{homeogrowth} as well as 
\eqref{eq:fbal} with \eqref{eq:def:stress:visc} we obtain} 
the dimensionless evolution equations
\begin{align}
  \label{eq:scaled:matter}
\DP{\rho}{t} + \DP{}{x}\left( \rho \DP{}{x}s \right) &= \left( 1 - \rho \right) \rho \\
\etascale \, \frac{\partial^2s}{\partial x^2} - s  &=   \log \rho + \param 
  \label{eq:scaled:const}
\end{align}
with boundary conditions 
\begin{subequations}
\label{eq:BC:scaled}
\begin{align}  
\label{eq:BC:scaled:0}
   \partial_xs (0, t) &= 0 \\
  \label{eq:BC:scaled:L:v}
  \partial_xs(L(t), t)   &= \dot{L}(t)\\
  \label{eq:BC:scaled:L:sigma}
  s(L(t), t) &= 0.
\end{align}
\end{subequations}
The dynamics of the free front depends only on \emph{two} 
dimensionless parameters: an active driving
\begin{equation}
  \label{eq:def:param}
  \param =  \frac{\sigp}{E}  + \log \frac{\rhoh}{\rhoe} 
\end{equation}
and a dimensionless viscous coefficient  
\begin{equation}
  \label{eq:def:etascale}
 \etascale = \frac{\eta}{E \, \td}.
\end{equation}

In order to discuss the physical origin of the active driving characterized 
by the parameter $\param$, it is instructive to consider the stationary, 
homogeneous solution of 
Eqs.~(\ref{eq:scaled:matter}-\ref{eq:BC:scaled}),
given by $\rho=1$, $s = -\param$, $v = 0$, and vanishing front velocity
$\dot{L}(t) = 0$. Returning to dimensionful quantities,
stationarity of the tissue requires that cell division and apoptosis 
balance on average, which according to Eq.~\eqref{homeogrowth} 
occurs for $\rho=\rhoh$. 
This state of tissue homeostasis implies that bulk stresses
are everywhere constant and equal to a homeostatic stress
\begin{equation*}
  \label{eq:def:sigh}
  \sigh = 
- E \log \frac{\rhoh}{\rhoe}  
\end{equation*}
which needs to be balanced at the tissue margin and thus requires 
$\sigp = \sigh$. Given the definition (\ref{eq:def:param}),
the stationary solution is only possible for zero active driving $\param = 0$.

The active driving $\param$ can thus be expressed as 
the difference between the protrusive stress exerted at the free edge
and the homeostatic stress, normalized by the tissue elastic modulus, 
\begin{equation*}
\label{eq:def:param:sigh}
\param = \frac{\sigp-\sigh}{E} \, . 
\end{equation*}
When $\param\neq0$, the homeostatic state cannot be sustained, and 
$\rho=\rho(x)\neq\rhoh$. In this case, net cell division 
($\kd>0$ for $\param>0$) or death ($\kd<0$ for $\param<0$) 
at the front give rise to tissue 
expansion or contraction, respectively, the dynamics of which is prescribed 
by Eqs.~(\ref{eq:scaled:matter}-\ref{eq:scaled:const}), 
see Sec.~\ref{sec:traveling:dynamics}.

A finite homeostatic stress $\sigh\neq0$ implies a mismatch between 
the carrying capacity $\rhoh$ and the elastic reference density $\rhoe$, which 
is often considered in elastic growth theories as a source of growth-induced 
stress (see \textrm{e.g.}~\cite{Rodriguez1994,Goriely2008}). The ratio between 
the applied stress at the free boundary to the elastic modulus of the tissue, 
$\sigp/E$, which quantifies the strength of the pulling forces exerted by the 
cells at the leading edge, can also be related to a spreading coefficient
within the context of wetting dynamics \cite{Douezan2012,Koepf2013}.

If $\sigp \gg E$, the tissue 
may no longer be able to accommodate large deformations 
and will eventually rupture under tensile stress \cite{Harris2012},
a phenomenon inconsistent with the present formulation which 
assumes continuity of the tissue. Conversely, if $-\sigp \gg E$, 
a buckling instability may occur: this eventuality is also beyond the
scope of the present quasi-1D approach. 
Recent work \cite{Podewitz2015,Harris2012} suggests values for 
$\sigh$ as well as for $E$ of the order of a few kPa.   
We therefore expect the active driving $|\param|$ to be at most of order $1$.

Before turning to the analysis, let us discuss the various scales
involved in this problem. For lack of measurements performed in 
the one-dimensional case, we rely on data from two-dimensional cell 
monolayers \cite{Puliafito2012,Harris2012,Cochet-Escartin2014}
as well as three-dimensional cellular spheroids
\cite{Guevorkian2010,Stirbat2013}.
The dependence of cell cycle duration upon cell number density
has been measured in \cite{Puliafito2012}, yielding a time scale
$\td \approx 10$ h $\approx 10^4$ s.
Using $E \approx 10^3 \,{\rm Pa}$ \cite{Harris2012} and
$\xi \approx 10^{16} \, \mathrm{Pa}\,\mathrm{m}^{-2}$ s 
\cite{Cochet-Escartin2014}, we deduce a length scale of the order of 
$\sqrt{E \td/\xi} \approx 30 \, \mu$m, and a velocity scale 
$U = \sqrt{E /(\xi\, \td)} \approx 10 \, \mu\mathrm{m} \, \mathrm{h}^{-1}$, 
similar to typical cell migration 
velocities~\cite{Vedula2012,Yevick2014,Yevick2015}.

In agreement with the typical viscosity of cell aggregates
\cite{Guevorkian2010,Stirbat2013}, the viscosity of a MDCK cell monolayer
has been measured in \cite{Harris2012}: from $\eta \approx 10^5\,{\rm Pa\,s}$, 
we find $\etascale \approx 10^{-2}$.
The associated viscous stresses can be estimated as follows. 
The velocity gradient in 
the tissue necessarily extends over several cells, and we estimate 
the order of magnitude of the strain rate as
$\del_x v \approx v_{\rm typ}/100\,\mu{\rm m} \approx 10^{-1} \,
\mathrm{h}^{-1}$ using the typical migration velocities 
mentioned above. Given a strain of order $\del_x u \approx 10^{-1}$ 
and with $E \approx 10^3 \,{\rm Pa}$ \cite{Harris2012},
$\eta \approx 10^5\,{\rm Pa\,s}$ \cite{Guevorkian2010,Stirbat2013}, 
one finds that $\eta \del_x v / E \del_x u \approx 2 \, 10^{-2}$.
One can \emph{a priori} expect viscous stresses to be negligible 
in the experiments. Note that most of the numerical values pertain 
to the epithelial MDCK cell line.

\section{Traveling waves}
\label{sec:traveling}

In this section, we first study numerically the system 
(\ref{eq:scaled:matter}-\ref{eq:BC:scaled}),
successively describing the transient dynamics 
(Sec.~\ref{sec:traveling:dynamics}) and the propagating front
(Sec.~\ref{sec:traveling:semiwave}) that follows in the asymptotic 
regime when the active driving is extensile. We then derive analytically 
an exact traveling wave solution in the limit of $\param \to \infty$ 
(Sec.~\ref{sec:traveling:smalldriving}).
The asymptotic front velocity is defined as 
$V=\lim_{t\rightarrow \infty}\dot{L}(t)$.

\begin{figure}[!t] 
\begin{center}
\showfigures{
\emph{(a)}
\includegraphics[scale=0.05]{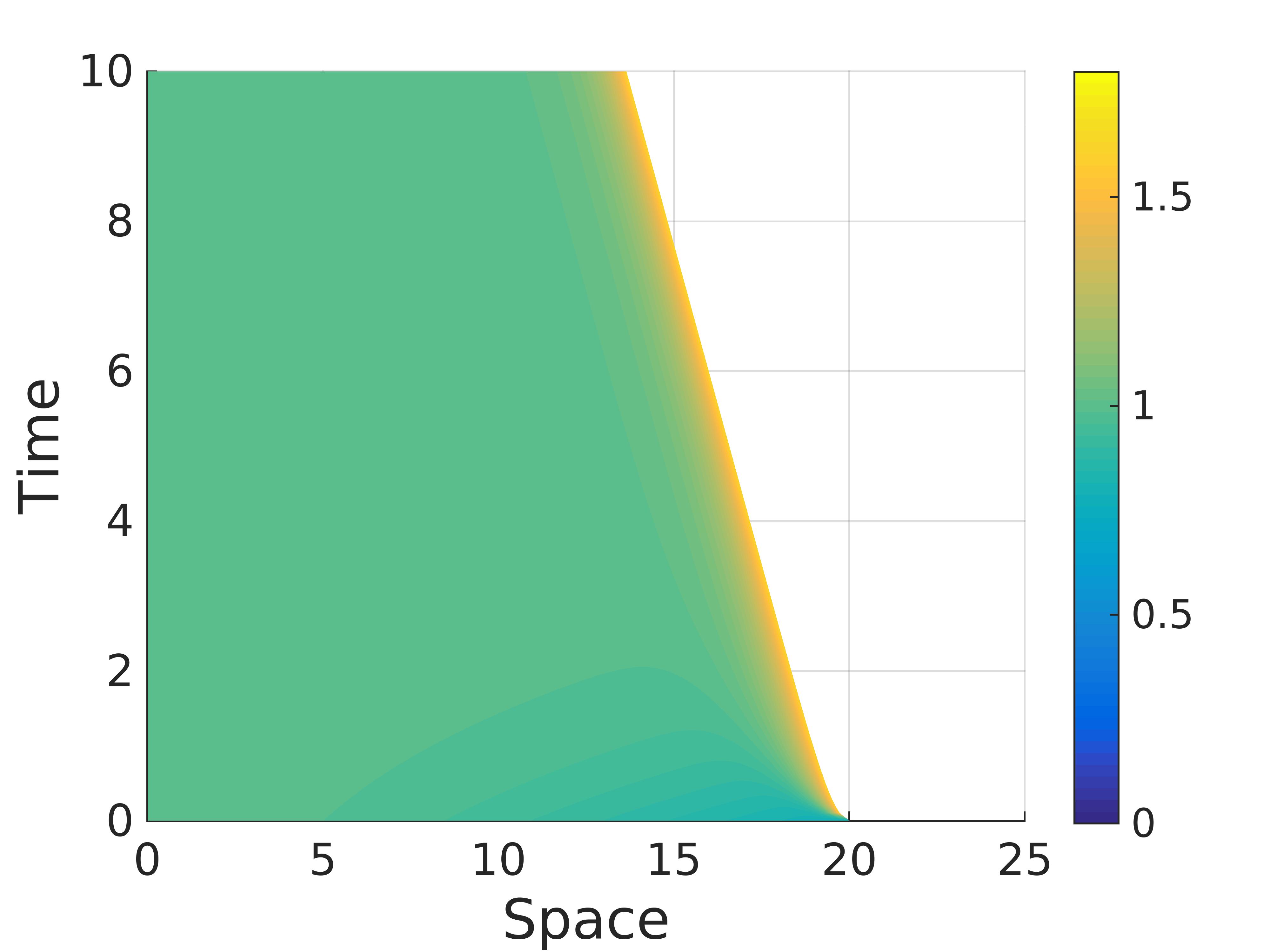}\\
\emph{(b)}
\includegraphics[scale=0.05]{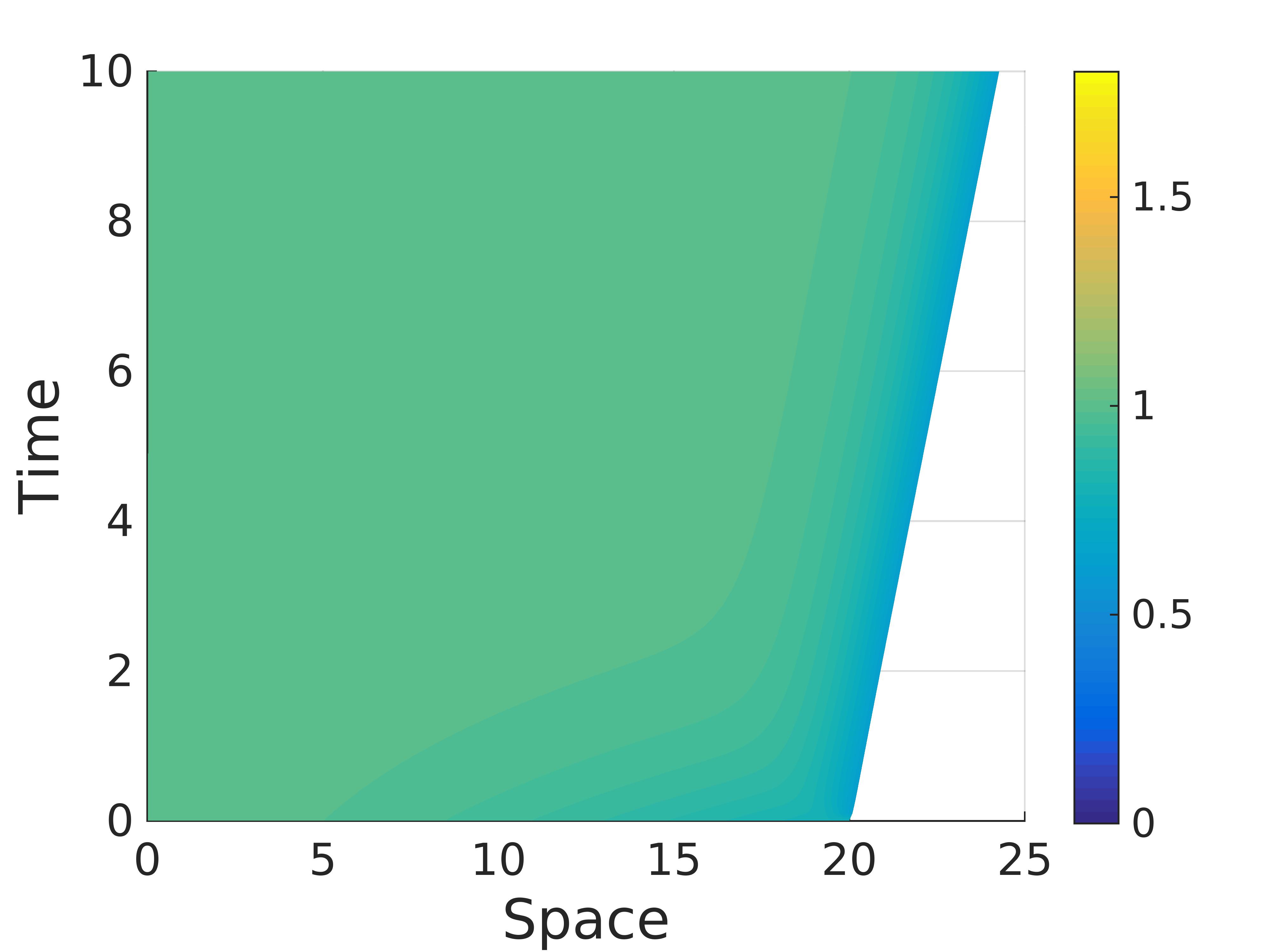}
}
\caption{
\label{fig_shrinking_spreading}
\textbf{Tissue contraction and expansion.}
Kymographs of the dimensionless cell density field $\rho(x,t)$ 
(density values given by the colourbar) showing 
\emph{(a)} contraction, $\param = -0.5$ 
and \emph{(b)} expansion, $\param = 0.5$. 
In both cases, $\etascale = 10^{-2}$ and
the initial condition for the cell density field is
$\rho^0(x) = \exp \left\{ -\left( x/(2L_0)\right)^2 \right\}$, 
$x \in [0, \, L_0]$, $L_0 = 20$.
(Online version in colour.)
}
\end{center}
\end{figure}

\subsection{Transients}
\label{sec:traveling:dynamics}

Numerical simulations of Eqs.~(\ref{eq:scaled:matter}-\ref{eq:scaled:const})
supplemented with boundary conditions (\ref{eq:BC:scaled})
are performed thanks to the numerical scheme described 
in Appendix~\ref{app:num}. 
Given the initial profile for the density field
$\rho(x,t=0) = \rho^0(x)$, $x \in [0,L_0]$, the initial profile $\sigma^0(x)$ 
for the stress field is obtained by solving (\ref{eq:scaled:const}), 
the spatial derivative of which gives the initial velocity field 
$v^0 = \mathrm{d}\sigma^0/\mathrm{d}x$.
We checked that the asymptotic behavior does not depend on the choice
of $\rho^0(x)$. As explained above, we qualitatively 
expect the monolayer to contract when $\param < 0$ and to expand 
when $\param > 0$, regardless of the value of the viscosity.

\paragraph*{Contraction.} 

Our simulations confirm that $\param<0$ leads to tissue contraction, see 
Fig.~\ref{fig_shrinking_spreading}(a) for a typical kymograph. 
Consistent with the qualitative argument presented in 
Sec.~\ref{sec:model:model}, one observes 
$\rho(x) \ge \rhoh$ close to the leading edge, 
implying a negative net cell division 
rate and progressive suppression of the tissue layer. 
After a short transient, contraction is approximately linear with time:
a propagative wave forms whose constant velocity $V$ is a function of the 
active driving $\param$ and the viscosity $\etascale$,
see Fig.~\ref{fig_vitesse}. 
One can define a characteristic time until collapse $t_{\rm c}$ as being 
the time when $L(t_{\rm c})=L_{\rm c}\ll1$. For large initial tissue sizes,  
$L(t=0)=L_0\gg1$, we find $t_{\mathrm{c}} \simeq L_0/V$.
Interestingly, recent numerical and experimental work \cite{Podewitz2015}
suggests that, in a number of cases, tissue homeostasis is a state 
of mechanical tension, characterized by a positive homeostatic stress 
$\sigh>0$. For zero protrusive stress $\sigp = 0$, Eq.~(\ref{eq:def:param})
then gives $\param < 0$: full inhibition of the protrusive
activity of leader cells may result in tissue contraction.

\paragraph*{Expansion.} 
In line with the experiments that 
motivate this work, we focus on the opposite case $\param>0$ in 
the remainder of this article. After a transient, whose duration 
increases with $\etascale$ and decreases with $\param$,
the tissue dynamics converges towards a spreading regime with 
a propagating front with constant velocity $V$, 
see Fig.~\ref{fig_shrinking_spreading}(b), in agreement
with observations \cite{Vedula2012,Yevick2015}. 
Here $\rho(x)\le\rhoh$: cell density decreases, whereas velocity 
increases with $x$, velocity being maximal at the leading edge.

\subsection{A propagating front} 
\label{sec:traveling:semiwave}

In the experimentally relevant case of monolayer expansion
$\param>0$, we next ask how the asymptotic front velocity $V$
depends on the parameters $\param$ and $\etascale$. 
From (\ref{eq:scaled:matter}-\ref{eq:BC:scaled}), it
satisfies the following problem on the half-axis 
$z=x-Vt\in ]-\infty,0]$, 
\begin{align}
\begin{split}
\label{eq:semiwave}
\begin{aligned}
-V\rho'+ (s' \rho)' &= \rho \, (1-\rho)\,,\\
\etascale  s'' - s &= \log \rho + \param \,,\\
\end{aligned}\\
\begin{aligned}
\rho(-\infty) &=1 \,, &s(0)&=0 \,, \\
s'(0)&=V\,, &s'(-\infty)&=0 \,, 
\end{aligned}
\end{split}
\end{align}
where $ \rho(z) = \rho(x,t)$, $ s(z) = s(x,t)$ and $z=0$ is 
the position of the free front. A prime $'=\DT{}{z}$ denotes the 
derivative with respect to the reduced variable $z$. 

\begin{figure}[!t] 
\begin{center}
\showfigures{
\includegraphics[scale=0.3]{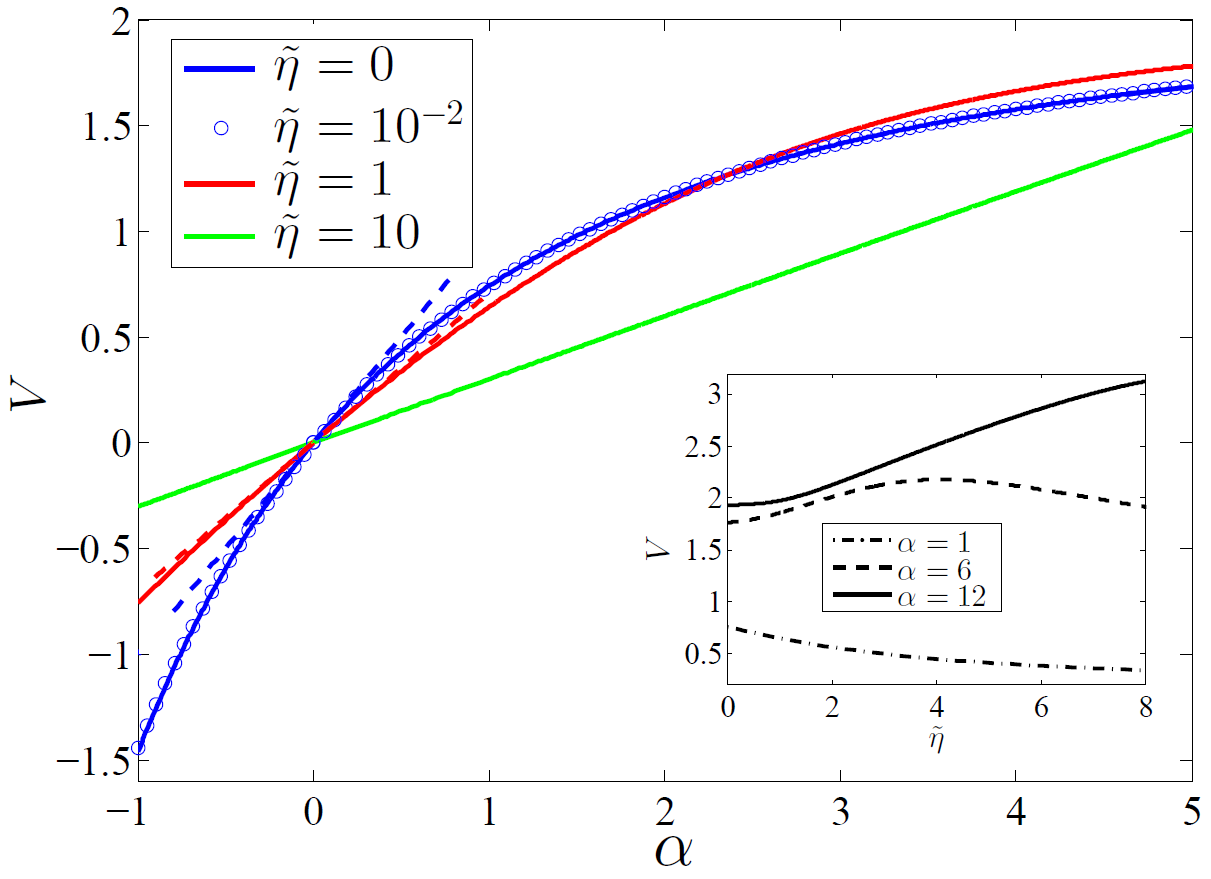}
}
\caption{
\label{fig_vitesse}
\textbf{Front velocity.} 
The dimensionless velocity $V(\param,\etascale)$ is plotted
as a function of the control parameter  $\param$, for several
values of the dimensionless viscosity $\etascale$.
Values obtained for $\etascale= 10^{-2}$ (blue circles) are indistinguishable 
from those obtained in the inviscid limit $\etascale = 0$ 
(solid blue line).
The dashed lines represent the linear approximation close to 
$(\param=0, V = 0)$: $V = \param/\sqrt{1+\etascale}$,
see Sec.~\ref{sec:traveling:smalldriving}. Inset: 
$V(\param,\etascale)$ \emph{vs.} $\etascale$ for fixed values 
of $\param$. For small enough $\param$ (dot-dashed curve) the velocity 
decreases with $\etascale$ as captured by linear analysis 
(see Sec.~\ref{sec:traveling:smalldriving}). For larger values of $\param$, 
note the non-monotonic viscosity-dependence of
the front velocity (dashed curve) which first increases with $\etascale$ 
and then decreases after a critical value of $\etascale$ is reached. 
This critical value increases with $\param$ and ultimately, 
when $\param\rightarrow \infty$, 
the front velocity becomes a monotonically 
increasing funtion of $\etascale$. (Online version in colour.)}
\end{center}
\end{figure}

To numerically compute $V(\param,\etascale)$, we choose to
operate by continuation from the known value $V=0$ at $\param=0$ 
using AUTO \cite{Doedel1981}. From this value, the software follows 
the solution of the nonlinear system \eqref{eq:semiwave} 
when $\param$ varies (negatively or positively) using a Newton algorithm. 
We checked that the velocities thus obtained are identical to those 
attained in the asymptotic regime using direct numerical simulation.
In Fig.~\ref{fig_vitesse}, we plot  $V(\param,\etascale)$ as a function
of $\param$ for several values of $\etascale$. Using the velocity scale 
$U = \sqrt{E /(\xi\, \td)} \approx 10 \, \mu\mathrm{m} \, \mathrm{h}^{-1}$
(see Sec.~\ref{sec:model:model}), the experimentally observed front velocities 
are of $\mathcal{O}(1)$ \cite{Vedula2012,Yevick2014} and thus suggest that 
$\param \approx 1-2$, consistent with our upper bound estimate
$\param \lessapprox \mathcal{O}(1)$.

\begin{figure}[!t] 
\begin{center}
\showfigures{
\includegraphics[scale=0.42]{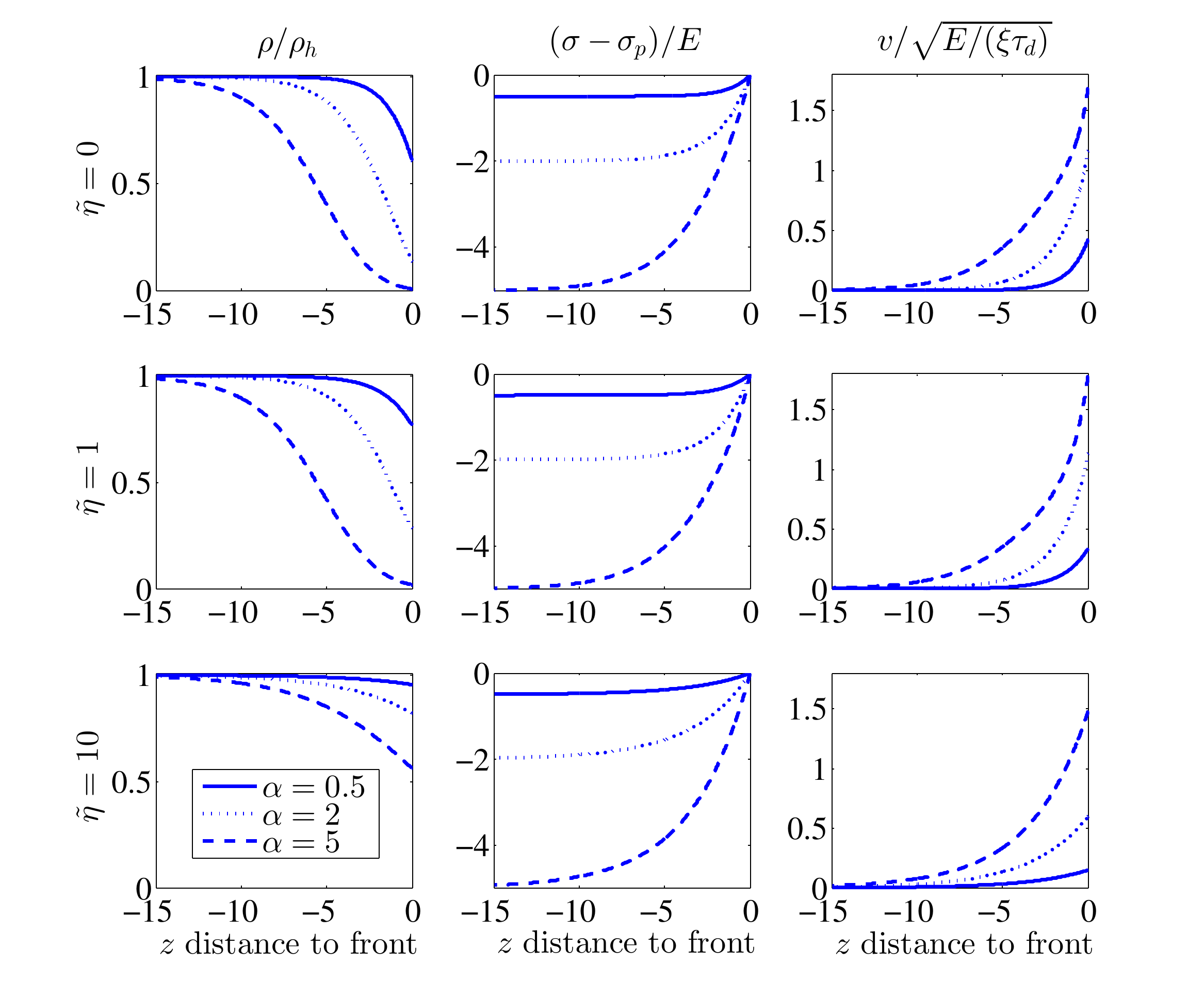}
}
\caption{
\label{fig_profile} 
\textbf{Typical steady-state profiles} of 
density, stress and velocity along the monolayer. 
(Online version in colour.)}
\end{center}
\end{figure}

We observe that the front velocity increases linearly with $\param$ for 
$|\param|\ll1$, and eventually saturates for $\param\to\infty$ to a value 
that \emph{increases} with $\etascale$.
The counterintuitive behaviour of $V$ with the viscosity 
can be understood qualitatively as follows. By integration of 
\eqref{eq:semiwave}, the front velocity can be expressed as a 
function of cell density only, 
\begin{equation*}
\label{eq:velocityint}
V=\int_{-\infty}^0\rho(z)(1-\rho(z)) \,\text{d}z.
\end{equation*}
Thus, cell proliferation contributes to propulsion only 
in the interfacial layer between the (homeostatic) state $\rho=1$ and 
the tissue margin density $\rho(z=0)$. The speed of the front depends both  
on the density range given by $1-\rho(0)$ 
and on the width $\lambda$ of the interfacial layer.
For $\param\to\infty$, the former is bounded by $1$, whereas $\lambda$ grows 
with $\etascale$ since viscous dissipation penalizes large gradients. 
As a consequence, in this limit, the front moves faster for a larger viscosity 
(see Fig.~\ref{fig_vitesse} inset, full line).
This situation is remindful of the ``pushed'' fronts observed 
in the propagation of an interface between two proliferating cell populations, 
where proliferation in the bulk occurs over a lengthscale that grows 
with tissue viscosity, propelling the interface forward~\cite{Ranft2014}. 
For $|\param|\ll1$, 
a perturbative calculation shows that the velocity grows with $\param$,
and decreases with $\etascale$ as $V = {\param}/{\sqrt{1+\etascale}}$ 
(see Sec.~\ref{sec:traveling:smalldriving}).
In this limit of small driving, the effect of $\etascale$ is 
two-fold: viscous dissipation both reduces $1-\rho(0)$ and decreases 
the slope of the interface region, the joint effect of which 
is a decrease in the front velocity (Fig.~\ref{fig_vitesse} inset, 
dotted-dashed line). In between these two limits of large and small driving, 
an increase of $\etascale$ may either increase or decrease the velocity 
depending on the values of both $\etascale$ and $\param$ 
(Fig.~\ref{fig_vitesse} inset, dashed line).

In Fig.~\ref{fig_profile}, we represent the steady-state profiles 
obtained by AUTO. Density, stress and velocity profiles connect 
the  homeostatic state $\rho = 1$, $s= -\alpha$, $v = 0$ 
in the bulk to the leading edge state $\rho=\rho(0)$, 
$s=0$, $v = V$ at $z = 0$. For all parameter values, the asymptotic 
density profiles decrease monotonically, whereas the velocity and stress 
profiles increase monotonically along the monolayer, in agreement with 
experimental observations. Whether the tissue is under 
tension ($\sigma>0$) or under compression (negative tension, $\sigma<0$) 
does not depend on the sign of $\param$ but on the signs of both 
$\sigh$ and $\sigp$. In general, 
$\min(\sigh,\sigp)\le\sigma\le\max(\sigh,\sigp)$, and while $s$ is 
always negative, the tissue can switch from a compressive to a tensile state 
at some bulk point (or even be tensile everywhere) due to the 
action of leader cells.

We checked numerically that the monotonic profiles of the cell density, 
velocity and stress fields found asymptotically are stable to perturbations
by an additive noise of small amplitude. 
Explaining the propagation of mechanical waves during 
two-dimensional tissue expansion \cite{Serra-Picamal2012}
will therefore require additional ingredients, among which 
contractility-dependent bulk motility forces are an obvious candidate.

\subsection{The small driving limit}
\label{sec:traveling:smalldriving}

For $\param\ll1$, 
the velocity of the expanding tissue can be explicitly found as a 
function of $\param$ and $\etascale$. Performing a Taylor expansion around the stationary state
$\param=0$, $s=0$, $\rho=1$, and $V=0$, we have
\begin{align*}
s &= 0+\epsilon\overset{1}{s}+o(\epsilon) \,, &\rho &= 1+\epsilon\overset{1}{\rho}+o(\epsilon) \,,\\
V &= 0+\epsilon\overset{1}{V}+o(\epsilon) \,, & \param &= 0+\epsilon\overset{1}{\param}+o(\epsilon) \,,
\end{align*}
where $\epsilon$ is a small, positive parameter $0 < \epsilon \ll 1$,
and $\overset{1}{q}$ denotes the first-order perturbation of quantity $q$. 
Eqs.~\eqref{eq:semiwave} then become at first order
\begin{align*}
\overset{1}{s}{}''=-\overset{1}{\rho} && \text{and} &&
\etascale \overset{1}{s}{}''-\overset{1}{s}=\overset{1}{\rho} + \overset{1}{\param}
\end{align*}
with boundary conditions
\begin{equation*}
\overset{1}{\rho}(-\infty) =
\overset{1}{s}{}'(0)-\overset{1}{V} = 
\overset{1}{s}{}'(-\infty) = \overset{1}{s}(0) = 0 \, .
\end{equation*}
Combining the first and second equation we obtain the second order 
differential equation
\begin{equation*}
(\etascale +1)\overset{1}{s}{}''-\overset{1}{s}=\overset{1}{\param} 
\end{equation*}
for the reduced stress field $s$ to first order.
Using boundary conditions on $\overset{1}{s}$ we obtain
\begin{align*}
\overset{1}{s} = \overset{1}{\param}
\left(e^{\frac{z}{\sqrt{1+\etascale}}} - 1 \right) 
&& \text{and} &&
\overset{1}{\rho} = -\overset{1}{\param}
\frac{e^{\frac{z}{\sqrt{1+\etascale}}}}{1+\etascale}\,.
\end{align*}
Spatial variations of velocity field, the stress field and  
the cell number density decay away from the front over 
a characteristic length 
\begin{equation*}
\label{eq:blayer:dimless}
\Delta z =\sqrt{1+\etascale} \, .
\end{equation*}
Since we expect $\etascale \simeq 10^{-2}$, the size of this boundary layer
is given by the unit of length, of the order of 
$\sqrt{E \td/\xi} \approx 30 \, \mu{\rm m}$. 

Of the two remaining boundary conditions, one is automatically satisfied 
while the last one provides the velocity,
\begin{equation}
\label{eq:blayer:velocity}
V = \frac{\param}{\sqrt{1+\etascale}} \, .
\end{equation}
This expansion for $\param\ll1$ can be made at all orders following the 
same procedure, defining an alternative way to analytically construct 
$V(\param,\etascale)$. We note that the linear approximation becomes accurate 
over a wider range of $\param$ as $\etascale$ increases, 
see Fig.~\ref{fig_vitesse}.

\section{The Inviscid limit}
\label{sec:inviscid}

In this section, we consider the inviscid limit of 
(\ref{eq:scaled:matter}-\ref{eq:scaled:const}) which corresponds to $\etascale\rightarrow 0$ 
and is more amenable to analysis. This limit can be viewed as 
the limit where the effective viscosity due to cell division $E\td$ 
dominates the bulk hydrodynamic viscosity of the tissue,
see Eq.~\eqref{eq:def:etascale}.
Plugging $s  = - \log \rho - \param$ (see \eqref{eq:scaled:const}) 
into \eqref{eq:scaled:matter} we obtain 
the following  parabolic reaction-diffusion equation for the cell density 
field $\rho$,
\begin{equation}
\label{eq:KPPS:evol:rho}
\partial_t\rho = \partial_{xx}\rho + \rho \, (1-\rho) \,,
\end{equation}
using dimensionless quantities.
When defined over the real axis, and for initial conditions decaying 
faster than exponentially, the Fisher-Kolmogorov 
equation~(\ref{eq:KPPS:evol:rho}) 
admits a traveling wave solution between the
fixed points $\lim_{x \to - \infty} \rho = 1$
and $\lim_{x \to + \infty} \rho = 0$ with a  velocity $V_{\mathrm{FK}} = 2 $ 
\cite{Fisher1937,Kolmogorov1937,Murray2002}. 
This equation is a classical model of collective cell migration
into empty space, originally introduced to describe the kinematics 
of wound healing assays \cite{Sherratt1990,Sherratt1992} by combining 
the effects of cell diffusion and cell proliferation, yet without 
reference to mechanical aspects.
Based on measurements of the front velocity and of the cell density profile,
good agreement has been found with predictions of the Fisher-Kolmogorov
equations for a variety of wound healing assays
\cite{Maini2004a,Maini2004b,Cai2007,Savla2004,Sengers2007,Simpson2013,Marel2014}. 
However, the smooth spatial variation of 
Fisher-Kolmogorov traveling waves is often hard to reconcile with 
the steepness of the cell density profile observed close to the 
leading edge \cite{Marel2014}. This point has led to the study 
of a sharp-front Fisher-Kolmogorov equation,
where the diffusion coefficient is a linear function of the cell density 
and vanishes when $\rho = 0$ \cite{Sanchezgarduno1995,Maini2004a,Sengers2007}.

Here, cell monolayer expansion corresponds to the associated free 
boundary problem posed on $x\in [0,L(t)]$  with the boundary conditions
\begin{subequations}
  \label{eq:KPP:BC}
\begin{align}
  \label{eq:KPP:BC:dx_rho:0:dimless}
\partial_x \rho(0, t) &= 0 \,, \\
  \label{eq:KPP:BC:rho:L:dimless}
\rho(L(t),t) &= e^{-\param} \,,\\
  \label{eq:KPP:BC:dx_rho:L:dimless}
\partial_x\rho(L(t),t) &=-\dot{L}(t) e^{-\param} \, .
\end{align}
\end{subequations}
Condition \eqref{eq:KPP:BC:rho:L:dimless} reflects the fact that the front is always sharp,
with a finite density that follows from Eq.~\eqref{eq:BC:sigma}, and 
diffusion has a purely mechanical origin, distinct from the random motion 
of single cells within the tissue.
Traveling wave solutions are known to exist for the problem 
(\ref{eq:KPPS:evol:rho}-\ref{eq:KPP:BC}) 
on the semi-axis $(-\infty, L(t)]$  \cite{Stepien2014} and
the result of Fisher-Kolmogorov is recovered in the limit of
strong active driving $\param\to \infty$. 

Indeed, taking the limit in the Dirichlet boundary condition,
Eq.~\eqref{eq:KPPS:evol:rho} is supplemented with
\begin{align*}
\begin{gathered}
\rho(L(t),t) = 0 \,, \\
\partial_x\rho(L(t),t)=0 \,, \quad \text{and} \quad \partial_x\rho(0,t)=0 \,.
\end{gathered}
\end{align*}
This problem belongs to the class
of models studied in \cite{bunting2012spreading}, where it is shown that 
regardless of initial conditions, linear expansion occurs. Further, 
Proposition 2.2 of 
\cite{bunting2012spreading} shows that the asymptotic velocity is
\begin{equation*}
  \label{eq:Phi0:V}
  \dot{L}(t)\underset{t\rightarrow \infty}{\longrightarrow} V_{\text{FK}}=2 
\end{equation*}
in agreement with the Fisher-Kolmogorov result.
We emphasize again that this limit of strong driving is unphysical. 

As in the viscous case, the dependence of the front velocity $V$
on the active driving $\param$ for the problem defined by 
Eqs.~(\ref{eq:KPPS:evol:rho}-\ref{eq:KPP:BC}) is obtained numerically
with the AUTO software \cite{Doedel1981}, see Fig.~\ref{fig_vitesse}.
The asymptotic front velocity $V(\param)$ is a monotonically 
increasing function interpolating between $V(0) = 0$ and $V(\infty)=2$.
A simple approximation giving the correct velocities and slopes
at both $\param=0$ and $\param=\infty$ is the function
$V_{\mathrm{approx}}(\param)=2 \left(1-e^{-\param/2} \right).$
As seen in Fig.~\ref{fig_vitesse}, the velocity curve 
$V(\param,\etascale)$ for a realistic viscosity $\etascale=10^{-2}$ is almost 
indistinguishable from $V(\param,0)$, consistent with our above  
reasoning (Sec.~\ref{sec:model:basis}) that viscous stresses are expected 
to be negligible.

In dimensional form, the front velocity is thus given by
\begin{equation}
\label{eq:KPP:velocity}
\mathcal{V} = \sqrt{\frac{E}{\xi \td}} \,
V\left( \frac{\sigp}{E} + \log\frac{\rhoh}{\rhoe} \right) 
\end{equation}
where the function $V$ is the dimensionless velocity computed numerically 
and drawn on Fig.~\ref{fig_vitesse}. 

We also show the asymptotic profiles for different values of $\param$, 
see Fig.~\ref{fig_profile}, top row. 
One can see that they display a boundary layer whose spatial extension
$\Delta x$ may be simply estimated in the following way,
\begin{equation*}
V(\param) = - \left( 
\frac{1}{\rho} \frac{\Delta \rho}{\Delta x} \right) \Big|_L 
\simeq  \frac{1-e^{-\param}}{e^{-\param}\Delta x}  \,,
\end{equation*}
or $\Delta x \simeq (e^{\param}-1)/V(\param)$. 
When $0< \param\ll 1$, in agreement with \eqref{eq:blayer:dimless} 
we recover $\Delta x=1$. When $\param \gg 1$, $\Delta x \to \infty$ 
since the transition between the values $\rho=1$ and $\rho=0$ 
may be located anywhere. Note however that the moving interface 
associated with the Fisher-Kolmogorov solution, where this transition occurs, 
still has a finite width of $\mathcal{O}(1)$.

\section{Possible effects of pharmacological perturbations} 
\label{sec:perturb}

For simplicity and as suggested by experimental data, we consider in 
this section only the inviscid limit, and discuss successively
the inhibition of cell division (Sec.~\ref{sec:disc:mitomycin}), 
of actin polymerization (Sec.~\ref{sec:disc:Rac}),  
and of contractility (Sec.~\ref{sec:disc:Blebb}).
In each case, Eq.~(\ref{eq:KPP:velocity}) allows in principle to predict the
response of the front velocity to pharmacological perturbations 
with drugs known to affect the cell cycle, the actomyosin cytoskeleton and/or 
to interfere with cell motility.

\subsection{Blocking cell proliferation}
\label{sec:disc:mitomycin}

When $\kd = 0$, the cell number density becomes a conserved 
quantity and the dimensional problem reads
\begin{align*}
\begin{gathered}
\label{eq:evol:nogrowth}
\partial_t\rho - \frac{E}{\xi} \partial_{xx}\rho = 0 \,, \\
\rho(L(t),t)=\rhoe \, e^{-\frac{\sigp}{E}} \,, \quad \partial_x\rho(0,t)=0 \,, \\
\dot{L}=-\frac{E}{\xi \rhoe} \, e^{\frac{\sigp}{E}} \, \partial_x\rho(L(t),t) \,.
\end{gathered}
\end{align*}
with an initial density profile $\rho(x,t=0) = \rho^0(x)$, 
$x \in [0,L_0]$. This is a classical Stefan problem for which (see Chap.~18 of 
\cite{Cannon1984}) the front will stop at the distance
\begin{equation}
  \label{eq:stop:length}
  L_{\mathrm{stop}} = e^{\frac{\sigp}{E}} \, 
\int_0^{L^0} \frac{\rho^0(x)}{\rhoe} \mathrm{d}x \,,
\end{equation}
which as expected increases with $\sigp$. 
When cell division is blocked by mitomycin during the collective migration 
of epithelial cells along cylindrical rods \cite{Yevick2015}, cells in the 
bulk stop moving, and monolayer expansion becomes confined to the front rows, 
where cells are stretched.
In addition, collective cell migration has been observed to stop at a finite
distance in the absence of cell division in a two-dimensional
scratch wound healing assay (see \cite{Mi2007}, where a finite 
$L_{\mathrm{stop}}$ was predicted on the basis of a Lagrangian description).
Both observations are in qualitative agreement with our model. 
However, whether the distance predicted by Eq.~(\ref{eq:stop:length}) 
is correct has not been tested quantitatively.

\subsection{Inhibiting actin polymerization}
\label{sec:disc:Rac}

Inhibitors of actin polymerization are expected to lower $\sigp$, and 
thus to lead to a lower front velocity through the decrease of $\param$, 
assuming all other parameters to be unchanged.
This was indeed observed experimentally on cylindrical wires of
radius $10 \, \mu$m using the Rac inhibitor NSC23766 \cite{Yevick2015}.

\begin{figure}[!t] 
\centering
\showfigures{
\includegraphics[scale=0.5]{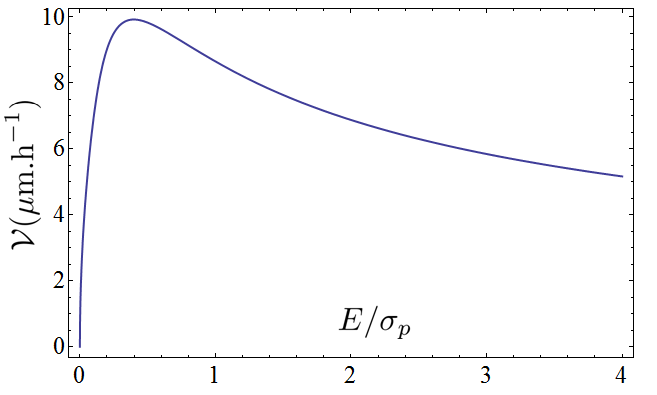}
}
\caption{\label{fig:contractility} 
\textbf{Dimensional front velocity $\mathcal{V}$ \emph{vs.}
effective elastic modulus $E/\sigp$},
for parameter values  $\sigh = 0$, $\td = 10^4$ s \cite{Puliafito2012},
$\sigp = 10^3 \text{ Pa}$, 
$\sigp/\xi=0.1 \, \mu\text{m}^2\text{s}^{-1}$ 
\cite{Cochet-Escartin2014}.}
\end{figure}

\subsection{Inhibiting contractility}
\label{sec:disc:Blebb}

Contractility may be taken into account explicitly in the model through
an additive, constant active stress $\siga > 0$ in the constitutive equation: 
$\sigma = - E \; \log( \rho/\rhoe) +\eta\partial_xv+ \siga$. 
Up to the definition of a modified protrusive stress 
$\sigp \to \tilde{\sigma}_{\mathrm{p}} = \sigp - \siga$, 
the problem is unchanged. Again assuming all other parameters to be 
unaffected, inhibiting tissue contractility decreases $\siga$,  
increases $\tilde{\sigma}_{\mathrm{p}}$, 
and therefore increases $\param$. This simple argument suggests that 
inhibiting contractility would lead to a higher front velocity. 

However, inhibitors of contractility may also modify tissue mechanics
through indirect (or non-linear) effects on parameters other
than $\siga$. In \cite{Cochet-Escartin2014}, some
among us conjectured that inhibiting the Rho pathway with
C3-transferase may also increase the friction coefficient $\xi$, 
due to alterations of the density and turn-over of 
cell-substrate adhesions. Here, the same effect would lead to a 
lower front velocity according to Eq.~(\ref{eq:KPP:velocity}).

Further, we expect  contractility inhibition to decrease the 
tissue elastic modulus, as blebbistatin treatment is known to soften 
cells \cite{Balland2005,Martens2008}.
In Fig.~\ref{fig:contractility},  we plot the velocity $\mathcal{V}$ of the 
moving front as a function of $E/\sigp$, as obtained from 
Eq.~(\ref{eq:KPP:velocity}), in the particular case 
$\rhoh = \rhoe$, \emph{i.e.} $\sigh = 0$. 
Strikingly, the curve $\mathcal{V}(E/\sigp)$ is not monotonic: 
a decrease of the tissue stiffness can result in either an increase 
or a decrease of the front velocity. Thus, the actual change of the front 
velocity may depend on the amplitude of the effect
of contractility inhibition on the elastic modulus.
Physically, this is due to the fact that in a softer material, growth 
generates less elastic stress pushing the free boundary but 
also resists less the pull generated by protrusive forces.

Together, we find that pharmacological inhibition of tissue contractility 
may therefore have an increasing or a decreasing effect on the front 
velocity depending on the concentration of the drug and
on the respective amplitude of its
impact on various physical parameters of the problem.
Experimentally \cite{Vedula2012,Yevick2015},  blebbistatin treatment
slows down the moving front in the case of narrow channels and fibers, 
whereas the effect is opposite for wider substrates.

Finally, let us emphasize that the possible effects of inhibitors have 
been inferred from known variations of the parameters
$\sigp$, $\siga$, $\xi$ and $E$. To our knowledge, possible 
effects on the parameters $\td$ and $\sigh$
have not been studied and may modify our conclusions.

\section{Conclusion}
\label{sec:conc}

A simple description of a cell monolayer as a one-dimensional,
proliferating, viscoelastic material allows
to reproduce qualitatively a number of experimental observations
pertaining to the expansion of epithelial monolayers 
in a laterally confined geometry \cite{Vedula2012,Yevick2015}: 
the displacement of the leading edge is linear in time; 
its velocity is of the order of $10 \,\mu \mathrm{m\,h}^{-1}$; 
the cell number density decreases while the velocity increases 
monotonically towards the moving front. 

The active control parameter combines two mechanisms: 
protrusive forces generated by active crawling at
the leading edge, and the mechanical effect of bulk cell proliferation. 
In the limit of small driving, an analytical solution
predicts exponential relaxation of the density, velocity and stress 
profiles over a short boundary layer.
In the limit of strong driving and zero viscosity, we recover
the Fisher-Kolmogorov equation, a classical model of collective
cell migration combining cell diffusion and proliferation. 
In this context, the diffusion coefficient 
receives a mechanical interpretation as the ratio of elastic modulus 
over tissue-substrate friction coefficient.
In the case of strongly cohesive tissues, such as epithelial
or endothelial monolayers, the cell density gradient is
steep at the front. Indeed our description postulates a finite cell density 
at the free boundary, due to the presence of actively migrating leader cells.
The Fisher-Kolmogorov approach predicts a smooth cell density gradient, 
and may thus better fit the expansion of high-density assemblies 
of mesenchymal cells, where cohesive forces are low and single
cell diffusion contributes to the collective behavior.

Our one-dimensional model may also describe some aspects
of the expansion of two-dimensional monolayers
\cite{Poujade2007,Trepat2009}, provided that translational invariance 
in the direction orthogonal to front spreading is a reasonable 
approximation, in a statistical sense, thus allowing for averaging.
Indeed similar one-dimensional models have been used
to describe aspects of two-dimensional tissue expansion
\cite{Maini2004a,Maini2004b,Mi2007,Arciero2011,Marel2014}.
While a proper tensorial generalization remains highly desirable, this
suggests that shear components may be neglected to first order when
describing the mechanics of cell monolayers.

The model depends on two dimensionless parameters,
the active driving $\param$ and the viscosity $\etascale$.
Although we provide an order of magnitude estimate for
$\etascale = 10^{-2}$, its relevance remains to be tested
quantitatively since it builds on measurements performed on
different geometries, sometimes with different cell types.
Ideally, one would like to fit the model to experimental data,
perhaps using velocity and cell density profiles to 
estimate model parameters. 
Given a linear relationship between proliferation rate and cell density, 
we used the framework of non-equilibrium linear thermodynamics and 
several simplifying, yet reasonable assumptions 
to predict that the tissue pressure should depend logarithmically 
on the cell density.
This prediction may be tested experimentally,
since the internal stress field can be obtained exactly in one
spatial dimension from traction force microscopy \cite{Trepat2009}.
As cell density increases in time, the same data may allow to estimate 
the critical value $\rhoe$ where tissue pressure changes sign.

We deliberately selected the minimal set of mechanical ingredients 
conducive to a constant velocity of tissue expansion, and thereby neglected, 
among other ingredients, bulk cell motility, cell polarity,
nonlinear tissue-substrate friction, or chemotaxis. 
It is our hope that appropriate modifications
may make this work relevant to the modeling of \emph{in vivo} 
collective cell migration \cite{Aman2010}, of which paradigmatic
examples are the formation of the lateral line primordium
\cite{Haas2006,Streichan2011}, or neural crest cells migration
\cite{Theveneau2012}.

\section*{Competing interests}
\label{sec:compint}

We have no competing interests.

\section*{Authors contributions}
\label{sec:auth:contrib}

P.R., J.R. and P.M. designed the study, performed research, and 
wrote the manuscript.
All authors gave final approval for publication. 

\acknowledgments

We wish to thank Hannah Yevick, Guillaume Duclos, 
Jean-Fran\c{c}ois Joanny, Pascal Silberzan and 
Lev Truskinovsky for the many enlightening discussions and suggestions 
that motivated this work, and Fran\c{c}ois Graner for a careful reading 
of the manuscript.

\appendix

\section{Evolution equation for arbitrary cell-division rate and elastic stress functions}
\label{sec:app:gen}

The above results are obtained for a specific form of the net cell division
rate $\kd(\rho)$, and an associated constitutive relation 
for the elastic stress that follows from non-equilibrium linear thermodynamics. 
For the sake of completeness, we give 
here the general evolution equations for the cell layer for arbitrary 
$\kd$ and elastic stress $\sige=-E \se(\rho)$, with the sole constraints that 
they decrease monotonically with the density, 
$\mathrm{d}\kd/\mathrm{d}\rho<0$, $\mathrm{d}\sige/\mathrm{d}\rho<0$, 
and vanish at finite values of the density.
We show that once these functions are 
fixed, the evolution still depends on only two parameters, namely the active 
driving $\param$ and the effective viscosity $\etascale$ identified above.

In the general case, the governing equations read
\begin{align}
\begin{gathered}
\label{app:general:gov_eqs}
\del_t \rho + \del_x (\rho v) = \kd(\rho) \rho \,,\\
\del_x \sigma = \xi v \,,\\
\sigma = -E \se(\rho) + \eta \del_x v \,,
\end{gathered}
\end{align}
 respectively expressing cell number balance, force balance and
the constitutive relation for the stress. They are supplemented 
with the boundary conditions
\begin{align*}
v(0) = 0 \,, && v(L)=\dot L \,, && \sigma(L) = \sigp \,.
\end{align*}

To simplify the notation in what follows, we introduce the following
(by now familiar) conventions:
\begin{equation*}
\begin{aligned}
\rhoh &\equiv \kd^{-1}(0) \,, \\
\rhoe &\equiv \se^{-1}(0) \,,\\
\sigh &\equiv - E \se(\rhoh) \,.
\end{aligned}
\end{equation*}
We furthermore define two auxiliary functions for the cell division rate
and the elastic stress relative to the homeostatic density, or carrying 
capacity, of the tissue:
\begin{equation*}
\begin{aligned}
\kappa(x) &\equiv \kd(x\rhoh) \,, \\
\sebar(x) &\equiv \se(x \rhoh) - \se(\rhoh) \,.
\end{aligned}
\end{equation*}
Non-dimensionalizing with the units 
\begin{equation*}
\begin{aligned}
t^* = 1/ \lim_{x\to0} \kappa(x) \,, &&
l^* = \sqrt{E t^* /\xi} \,, &&
\rho^* = \rhoh 
\end{aligned}
\end{equation*}
for time, length, and cell number density, respectively, and
using the previously introduced reduced stress
\begin{equation*}
s= \frac{\sigma-\sigma_p}{E} \,,
\end{equation*}
we can then recast Eqs.~\eqref{app:general:gov_eqs} in the simpler form
\begin{align}
\label{app:general:eqs_dimless}
\begin{gathered}
\DP{\rho}{t} + \DP{}{x}\left( \rho \DP{}{x}s \right) = \kbar(\rho)\rho \,,\\
\etascale \, \frac{\partial^2s}{\partial x^2} - s =   
\sebar(\rho) + \param \,,
\end{gathered}
\end{align}
where $\kbar(\rho)=\kappa(\rho)/{\lim_{x\to0} \kappa(x)}$.
The boundary conditions correspondingly become
\begin{align*}  
\partial_xs (0) = 0 \,, &&  \partial_xs(L)  = \dot{L} \,, && s(L) = 0 \,.
\end{align*}
As in the specific case discussed in the main manuscript, for any given 
functions $\kd$ and $f$ the dynamics depends only on two dimensionless
parameters given by
\begin{align*}
\param = \frac{\sigp - \sigh}{E} \,, &&
\etascale = \frac{\eta}{E t^*}\,.
\end{align*}
Stationarity of~\eqref{app:general:eqs_dimless} requires $\rho=1$, $s=-\param$, 
and can only be attained for $\param=0$ when taking the boundary condition 
on $s$ into account.

We checked numerically that linear expansion is also observed for 
several plausible choices of the function $\se(\rho)$
(see \cite{Stepien2014} for a proof in the inviscid limit).
The curve $V(\param,\etascale)$ 
depends quantitatively on the precise form of the equation of state.

\section{Numerical resolution of the free boundary, viscous problem}
\label{app:num}

\subsection{Scaled variables}
\label{sec:num:scaled}

For the numerical resolution, in order to write boundary conditions at 
a fixed position in space, we prefer the scaled coordinate
\begin{equation}
  \label{eq:def:y}
  y = \frac{x}{L(t)}.
\end{equation}
and denote the new unknown functions 
$\hat{\sigma}(y,t)=\sigma[L(t)y,t]$ and 
$\hat{\rho}(y,t)=L(t) \, \rho[L(t)y,t]$.
Eq.~(\ref{eq:cons:matter}) becomes
\begin{equation}
  \label{eq:evol:matter}
\DP{\hat{\rho}}{t}
 +\frac{1}{L}  \DP{}{y}\left(\hat{v}\hat{\rho}\right) =
\hat{\rho}\left( 1 - \frac{\hat{\rho}}{L} \right)
\end{equation}
where the velocity field  relative to the leading edge velocity can be expressed through the momentum conservation equation
\begin{equation}
  \label{eq:evol:momentum}
\frac{1}{L} \DP{\hat{s}}{y}-y\dot{L} = \hat{v}
\end{equation}
Using (\ref{eq:evol:momentum}), 
the constitutive equation (\ref{eq:scaled:const}) becomes
\begin{equation}
  \label{eq:evol:const}
 \frac{\etascale}{L^2} \; \DPn{2}{\hat{s}}{y}-  \hat{s} =  \log(\frac{\hat{\rho}}{L}) +\param
\end{equation}
Accordingly, the boundary conditions (\ref{eq:BC:scaled}) become
\begin{subequations}
\label{eq:BC:evol}
\begin{align}  
  \label{eq:BC:evol:0}
  \hat{v}(y=0, t) &= 0 
\\
  \label{eq:BC:evol:1:v}
  \hat{v}(y = 1, t)   &= 0
\\
  \label{eq:BC:evol:1:sigma}
  \hat{s}(y = 1, t) &= 0
\end{align}
\end{subequations}

\subsection{Numerical implementation}
\label{sec:num:scheme}
The numerical scheme used to solve the Cauchy problem 
Eqs.~(\ref{eq:evol:matter}-\ref{eq:BC:evol}) 
is based on the finite volume method \cite{Leveque2002} in order 
to strictly conserve mass and handle very localized states without 
spurious oscillations.  

Two regularly-spaced grids on the same interval $[0,1]$, 
denoted $Z$ and $Z_d$ for its dual, are considered in parallel.
An initial condition on $\hat{\rho}$ being given on $Z$, 
\eqref{eq:evol:const} is solved using boundary conditions 
\eqref{eq:BC:evol:0} and \eqref{eq:BC:evol:1:sigma} and the effective drift 
term $\hat{v}$ is computed on $Z_d$ using relation \eqref{eq:evol:momentum}. 
We then apply an upwind finite volume scheme to \eqref{eq:evol:matter} using 
the no flux boundary conditions \eqref{eq:BC:evol:0} and \eqref{eq:BC:evol:1:v}.
This allows the computation of the updated concentration profile 
$\hat{\rho}$ on $Z$ which gives in turn the new initial
data used for the next time step. The same procedure is then repeated. 
The time interval for each time step is
adapted so that the Courant-Friedrichs-Lewy condition is uniformly 
satisfied on $Z_d$ \cite{Leveque2002}.

\bibliographystyle{rsc}
\bibliography{migration}


\end{document}